\documentclass[10pt]{article}
\usepackage[utf8]{inputenc}
\usepackage{amsmath}
\usepackage{amsfonts}
\usepackage{amssymb}
\usepackage{authblk}

\usepackage{kbordermatrix}
\usepackage[all]{xy}
\usepackage{blkarray}
\usepackage{multicol} 
\usepackage{booktabs}
\usepackage{multicol}
\usepackage{setspace} 
\usepackage{multicol}
\usepackage{orcidlink}

\usepackage{graphicx}

\usepackage{ulem} 
\usepackage{pdflscape}
\usepackage{hyperref}

\usepackage[ruled,vlined]{algorithm2e}
\DontPrintSemicolon   

\usepackage[left=2.5cm, right=2.5cm, top=2.0cm, bottom=2.5cm]{geometry}

\title{{\LARGE\textbf{Unveiling Cancer Stem Cell Marker Networks:\\ A Hypergraph Approach}}}

\begin{scriptsize}

\author[1,2,\orcidlink{0000-0003-1946-0413}]{David H. Margarit}
\author[3,5,\orcidlink{0000-0002-0895-1957}]{Gustavo Paccosi}
\author[1,2,4,\orcidlink{0000-0002-9856-7501}]{Marcela V. Reale}
\author[1,2,\orcidlink{0000-0002-9272-6575}]{Lilia Romanelli}

\affil[1]{Instituto de Ciencias (ICI), Universidad Nacional de General Sarmiento (UNGS). J. M. Guti\'errez 1150, Los Polvorines, B1613, Buenos Aires, Argentina.}
\affil[2]{Consejo Nacional de Investigaciones Cient\'ificas y T\'ecnicas (CONICET). Godoy Cruz 2290,
Ciudad Aut\'onoma de Buenos Aires, C1425, Argentina.}
\affil[3]{Instituto de Desarrollo Humano (IDH), Universidad Nacional de General Sarmiento (UNGS). J. M. Gutirrez 1150, Los Polvorines, B1613, Buenos Aires, Argentina.}
\affil[4]{Departamento de Ingenier\'ia e Investigaciones Tecnol\'ogicas (DIIT), Universidad Nacional de la Matanza (UNLaM). Florencio Varela 1903, San Justo, La Matanza, B1754, Buenos Aires, Argentina.}

\end{scriptsize}

\date{}
\begin{document}

\maketitle

\begin{abstract}
We propose a novel computational framework leveraging hypergraph theory to analyse cancer stem cell markers (CSCMs) across multiple organs. Hypergraphs provide a robust representation of CSCM co-expression patterns, capturing their complex multi-organ relationships more comprehensively than traditional graph-based methods. By integrating mutual information analysis and Markov models, we identify key markers driving tumour heterogeneity and metastasis, offering detailed insights into their interdependencies. This approach establishes hypergraphs as a computationally powerful tool to model cancer progression and metastatic dynamics, contributing to the understanding of complex biological systems and supporting the development of targeted therapeutic strategies.


\end{abstract}

\subsection*{Keywords}
Hypergraphs, Biological Modelling, Cancer Stem Cell Markers, Cancer Progression

\subsection*{MSC Classification}
92Bxx, 05Cxx, 60J05, 92-10.

\section{Introduction}

Cancer is a group of diseases characterised by the uncontrolled growth of abnormal cells, which can invade nearby tissues and spread to other organs \cite{cancer1, cancer2}. Genetic mutations allow these cells to evade apoptosis and bypass the immune system. As tumours grow, they disrupt tissue function and may trigger angiogenesis, providing nutrients and oxygen through newly formed blood vessels. If untreated, cancer cells can disseminate via the bloodstream or lymphatic system, leading to metastasis \cite{cancer2}.

Among malignant tumour cells, cancer stem cells (CSCs) stand out due to their capacity for self-renewal and differentiation into multiple cell types \cite{stem}. These cells are pivotal in tumour initiation, progression, and resistance to standard therapies, contributing to tumour heterogeneity \cite{d, e}. CSCs are identified by cancer stem cell markers (CSCMs), such as $CD44+$ and $CD133+$, whose expression varies by cancer type \cite{a}. A single CSCM may be present in multiple organs or tissues, complicating their characterisation \cite{b}. Understanding their distribution is essential for developing targeted therapies, particularly in the context of metastasis.

Traditional cancer modelling approaches have proven valuable for simulating tumour growth and predicting treatment outcomes \cite{camodel1, camodel2}. However, many of these rely on simplified network representations that fail to capture the full complexity of CSCM distributions across organs. In contrast, hypergraph theory provides a computational framework that models multi-node connections through hyperedges, where each CSCM links all the organs in which it is expressed \cite{battiston}. This approach not only preserves the complexity of co-expression but also supports quantitative analyses with metrics such as vertex and hyperedge degrees.

Further, mutual information and Markov chains enhance this computational framework. Mutual information quantifies interdependencies between organs and CSCMs, revealing key markers driving tumour heterogeneity and metastasis \cite{rosas}. Markov chains simulate the probabilistic dynamics of CSC-mediated metastatic pathways, characterising how CSC presence in one organ increases colonisation likelihood in others.

This work presents a computationally grounded framework combining hypergraph theory, mutual information, and Markov chains to study CSCM distributions and metastatic dynamics. This integrative approach advances the computational modelling of cancer systems, bridging biological complexity and theoretical insights for broader applications in cancer research.

\section{General objectives and methodology\label{metodologia}}

This study utilises the comprehensive dataset compiled by \cite{yang2020}, which integrates data from various databases and reviews regarding the distribution of surface CSCMs (cellular markers found on the walls of CSCs) across different cancer-affected tissues or organs. This dataset forms the foundation for our hypergraph-based analysis, aimed at identifying CSCMs distribution patterns across organs. The hypergraph approach proves advantageous as it captures co-expression relationships that traditional graph models cannot represent. 

In our model, each cancer-affected organ is represented as a node, while CSCMs are represented as hyperedges connecting these nodes. We construct an incidence matrix $H=(V,E)$ \cite{dai2023}, where $V$ denotes the set of nodes (organs), and $E$ denotes the set of hyperedges (CSCMs). Formally, each element $H_{ij}$ of the matrix is defined as 1 if the organ $v_i$ is associated with the CSCM $e_j$, and 0 otherwise:

\begin{equation}
H_{ij}=h(v_i,e_j) = \left \{
\begin{array}{l l}
1 & \mbox{if } v_i \in e_j \\ 
0 & \mbox{if } v_i \notin e_j 
\end{array}
\right.
\label{defH}
\end{equation}

\begin{figure}
\begin{center}
\includegraphics[width=1\linewidth]{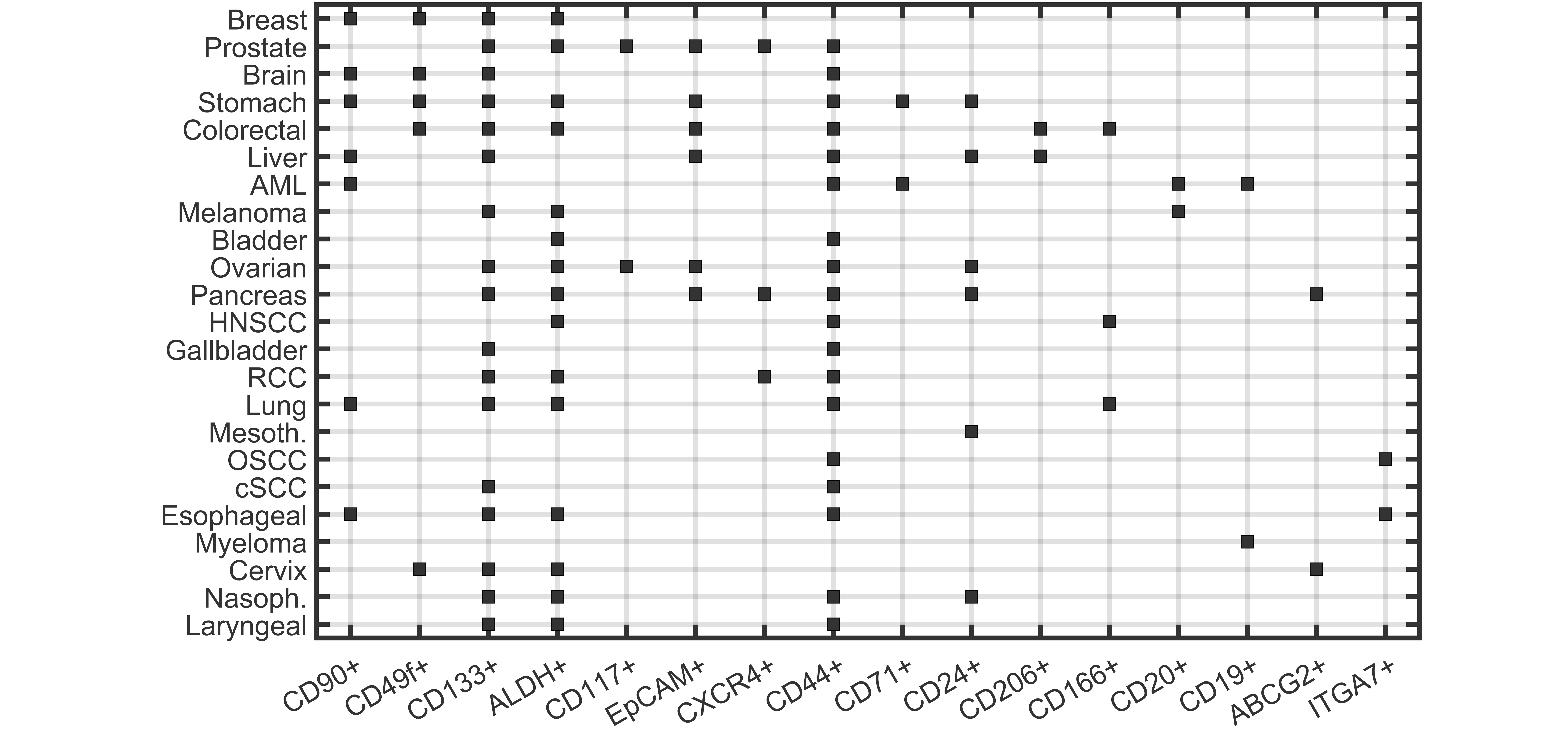}
\caption{\label{fig:incidencia} Representation of the incidence matrix $H$. Each square indicates the node-hyperedge (organ-CSCM) interconnection.}
\end{center}
\end{figure}

We analyse 23 organs and 16 CSCMs, constructing an incidence matrix $ H $ with dimensions $ 23 \times 16 $, as shown visually in Fig. (\ref{fig:incidencia}). This analysis provides valuable insights into the biological roles of CSCMs in tumour heterogeneity and metastasis, offering potential strategies for targeting a wide range of cancer types.


The sets of vertices $V$ (organs) and hyperedges $E$ (CSCMs) are defined as follows:

 $ V= \!\{\! v_1,\dots,v_{23}\! \}\! =\!\{$\!\textit{breast, prostate, brain, stomach, colorectal, liver, AML, melanoma, bladder, Ovarian, Pancreas, HNSCC, gallbladder, RCC, lung, mesothelioma, OSCC, cSCC, esophageal, myeloma, cervix, nasopharyngeal, laryngeal}$\}$.

 $E=\{e_1, \dots , e_{16}  \} =\{$\textit{CD90+, CD49f+, CD133+, ALDH+, CD117+, EpCAM+, CXCR4+, CD44+, CD71+, CD24+, CD206+, CD166+, CD20+, CD19+, ABCG2+}$\}$

For clarity, some cancer types are abbreviated as follows: for acute myeloid leukaemia (\textit{AML}), head and neck squamous cell carcinoma (\textit{HNSCC}), renal cell carcinoma (\textit{RCC}), oral squamous cell carcinoma (\textit{OSCC}), and cutaneous squamous cell carcinoma (\textit{cSCC}).
 
\subsection{Brief characterization of hypergraph $H$}
The incidence matrix $H$ offers a structured representation of connections between cancer-affected organs (nodes) and CSCMs (hyperedges) in the hypergraph. The degrees of vertices (organs) and hyperedges (CSCMs) provide essential metrics to understand the connectivity and centrality within the system. These degrees quantify how extensively an organ is associated with CSCMs and how broadly a CSCM is distributed across organs.

In the hypergraph framework, the degree of a vertex $v$ (organ) is defined as:
$$ d(v)=\sum_{e \in E} h(v,e)= \vert \mathcal{E}(v)\vert ,$$
where $$\mathcal{E}(v) = \{ e \in E: v \in V \}.$$ 
Similarly, the degree of a hyperedge $e$ (CSCM) is given by:
$$ \delta(e)=\sum_{v \in V} h(v,e)=\vert e \vert$$

Extending this concept to a matrix form, $D_v$ and $D_e$ represent the diagonal matrices of the vertex and hyperedge degrees, respectively, as shown in Eq. (\ref{Dv}) and Eq. (\ref{De}). These values help evaluate the centrality of organs and the distribution of CSCMs, offering insights into tumour heterogeneity and the role of specific markers in cancer progression  \cite{dai2023,book1}.

\begin{equation}
D_v=diag(D_{v_{1,1}},\dots,D_{v_{23,23}})=(4,6,4,8,7,6,5,3,2,6,7,3,2,4,5,1,2,2,5,1,4,4,3)
\label{Dv}
\end{equation}

\begin{equation}
D_e=diag(D_{e_{1,1}},\dots,D_{e_{16,16}})=(7,5,17,15,2,6,3,18,2,6,2,3,2,2,2,2)
\label{De}
\end{equation}

This analysis uncovers key patterns in tumour dynamics: organs such as the \textit{stomach}, with a degree of 8, are strongly connected to multiple CSCMs, while markers like $CD133+$, $CD44+$, and $ALDH+$ exhibit high degrees, linking to various organs. These degree definitions are intrinsic to the hypergraph and do not require a bipartite representation. They quantify the direct associations between nodes and hyperedges, offering a quantitative perspective on the system's connectivity. By integrating degree metrics with mutual information, the analysis delves deeper into how specific CSCMs in particular organs influence tumour behaviour. This combined approach elucidates both the structural organisation of the hypergraph and the functional relationships between CSCMs and organs, shedding light on cancer progression mechanisms.

The order ($\mathcal{O}_H$) of the hypergraph is the number of vertices, which in this case is 23:
$$\mathcal{O}_H=23.$$
The size ($\mathcal{T}_H$) is the number of hyperedges, which corresponds to 16 CSCMs:
$$ \mathcal{T}_H=16.$$
The rank ($\mathcal{R}_H$) and co-rank ($co\mathcal{R}_H$) of the hypergraph are determined by the sizes of the largest and smallest hyperedges:
$$ \mathcal{R}_H=17 \qquad \mbox{and} \qquad  co\mathcal{R}_H=2.$$
These metrics highlight the diversity of interactions, as the hypergraph is not \textit{r-uniform} (a term referring to hypergraphs where hyperedges have uniform sizes). The variation in hyperedge sizes reflects the heterogeneity in the system, where certain CSCMs are linked to many organs, whereas others are more specific to certain organs. This non-uniformity is crucial for understanding the heterogeneous nature of cancer and how specific CSCMs might drive metastasis in different ways  \cite{book1}. Given the complexity of the matrix, a detailed visualisation is provided in Fig. (\ref{fig:incidencia}) to aid in understanding, while further details about $H$ are included in Appendix \ref{Hmatrix}.

The sequence of computational steps used to analyse this work and its associated information-theoretic measures is summarised in Algorithm~\ref{algor}, located in Appendix~\ref{algoritmo}. 

\section{From hypergraph structures to Markovian processes\label{markovianos}}
  Metastasis arises from stochastic factors such as genetic mutations, tumour plasticity, and interactions within the tumour microenvironment \cite{majid}. In this study, we employ Markov chains to model transitions between cancer-affected organs, using the distribution of CSCMs as a proxy for metastatic potential.

Shared CSCMs reflect common microenvironmental conditions that enable CSCs to survive and colonise distant organs, as demonstrated by the metastatic tropism associated with markers like $EpCAM+$ and $CD44+$ \cite{majid}. These markers serve as proxies for CSC activity and plasticity, which influence the stochastic migration of CSCs through the bloodstream or lymphatic system. While CSCMs themselves do not migrate, their distribution across organs provides valuable insights into the migration and colonisation patterns of CSCs, which drive metastasis by circulating and establishing in distant tissues. The transition probabilities in the Markov model are derived from the co-expression levels of CSCMs, revealing potential metastatic pathways. This approach is supported by evidence that CSCs are more likely to migrate between organs with shared markers, due to similarities in their tumour microenvironments \cite{cancer2}.

Traditional random walks on simple networks capture pairwise interactions but fail to account for the complex, multi-organ dynamics mediated by CSCs. In contrast, random walks on hypergraphs, as discussed by \cite{carletti}, enable the Markovian framework to model the probabilistic migration of CSCs between organs. This approach aligns with biological evidence suggesting that CSCs migrate via pathways such as the bloodstream or lymphatic system to invade new tissues. By incorporating shared CSCMs into the transition probabilities, we can identify potential metastatic routes and highlight organs that may act as hubs in this process.

This framework conceptualises cancer spread as a probabilistic process driven by CSC migration rather than the direct propagation of CSCMs. Hypergraphs overcome the limitations of traditional random walks and pairwise networks by capturing higher-order interactions between organs and CSCMs. In this model, the degree of a hyperedge (representing a CSCM) indicates its involvement in multiple organs, providing a more accurate representation of the pathways through which CSCs may migrate. Higher-degree CSCMs, which are present in multiple organs, suggest a higher probability of CSC migration across these organs, thereby helping to identify key routes in cancer spread.

Markov chains facilitate a detailed analysis of transition matrices derived from hypergraphs, modelling the dynamic relationships between nodes (organs) and hyperedges (CSCMs) through transition probabilities \cite{markov}. These matrices illustrate how CSCs, indicated by shared CSCMs, may migrate between organs over time, revealing key patterns in cancer spread. For instance, transition probabilities quantify the likelihood that an organ becomes influenced by CSCs associated with a specific marker, or that a marker is shared across multiple organs, providing critical insights into metastatic pathways. The transition probabilities derived from shared CSCMs reveal potential metastatic routes. For example, high probabilities between the stomach and liver align with clinical observations of frequent metastases in gastrointestinal cancers \cite{gastro}. This framework not only identifies potential pathways but also highlights organs likely to act as hubs in metastasis, aiding in the prioritisation of therapeutic targets.

\subsection{Dynamic Modeling from Hypergraph}

To implement the concepts discussed earlier and interpret the findings, we use the framework of a \textit{random walk} on hypergraphs. In this model, a \textit{lazy random walker} moves from node $u$ at time $t$ to another node $v$ at time $t+1$. The random walk on a hypergraph consists of two steps: first, the walker randomly selects a hyperedge $e$ connected to the current organ (vertex) $u$; second, a new organ $v$ is randomly chosen from within the selected hyperedge $e$ \cite{eriksson}. While cancer spread is not entirely random, the random walk on hypergraphs effectively captures the probabilistic influence of CSCMs across multiple organs. This methodology offers a useful framework for simulating the development of metastatic pathways in uncertain environments.

As a reminder, the degree of a vertex (organ) $d(u)$ and the degree of a hyperedge (CSCM) $\delta(e)$ in an unweighted hypergraph are defined as:

$$ d(u) = \vert \mathcal{E}(u) \vert \quad \text{and} \quad \delta(e) = \vert e \vert $$

where $\mathcal{E}(u)$ is the set of hyperedges incident to $u$.

To facilitate this reasoning, we focus on two nodes, $u$ and $v$, and a single hyperedge $e$ connecting them. The first step involves moving from node $u$ along hyperedge $e$, quantified by the probability:
$$ p_{ue} = \frac{1}{d(u)}$$
The second step involves transitioning from hyperedge $e$ to node $v$, quantified by the probability:
$$ p_{ev} = \frac{1}{\delta(e)}$$

From $p_{ve}$, we can generalise the transition matrix $P_{ve}$, which represents the probabilities of moving from an organ (vertex) to a CSCM (hyperedge) for all organs and markers, capturing the dynamics of these interactions \cite{quantum,dai2023}.

\begin{equation}
P_{ue} = D_v^{-1}H
\label{pve}
\end{equation}

Similarly, the transition probability $ p_{ev} $ from a CSCM (hyperedge) to a specific organ (vertex) is captured by the transition matrix $ P_{ev} $, defined as:

\begin{equation}
P_{ev} = D_e^{-1}H^T
\label{pev}
\end{equation}

These matrices satisfy the following normalisation properties:

$$\sum_{e \in E} p_{ue} = 1 \quad \forall\, v \in V \quad \text{and} \quad \sum_{v \in V} p_{ev} = 1 \quad \forall\, e \in E$$

This normalisation ensures that the transition probabilities in each row sum to 1, which is essential for analysing Markovian systems where the total probability must be conserved \cite{markov}.

Using these transition matrices, we can analyse the probabilistic paths through which CSCs spread across organs. This approach reveals both direct and indirect relationships between organs via shared CSCMs, which serve as indicators of potential metastatic routes. For example, transitions between organs may occur through CSCs characterised by shared markers, providing a clearer picture of metastatic pathways and their underlying mechanisms. While cancer spread is not entirely random, the random walk on hypergraphs models the probabilistic nature of how CSCs, marked by specific CSCMs, influence multiple organs, simulating the development of metastatic routes in uncertain environments.

Finally, summarising the previous steps, the transition probability $ p(u, v) $ represents the likelihood that the chain moves from organ $u$ to organ $v$ at the next time step \cite{Bellaachia}:

$$ p(u, v) =  \sum_{e \in \mathcal{E}(u) \cap \mathcal{E}(v)} \frac{1}{d(u)\delta(e)} = \frac{1}{d(u)} \sum_{e \in \mathcal{E}(u) \cap \mathcal{E}(v)} \frac{1}{\delta(e)} $$

Alternatively, extending this formulation to matrix form, we compute the vertex-vertex transition matrix $ P_{uv} $ via the expression:

\begin{equation}
P_{uv} = P_{ue} P_{ev},
\label{pvv}
\end{equation}

where $ P_{ue} $ and $ P_{ev} $ represent the organ-CSCM and CSCM-organ transitions, respectively, as defined by Eqs. (\ref{pve}) and (\ref{pev}).

\begin{figure}
\begin{center}
\includegraphics[width=1\linewidth]{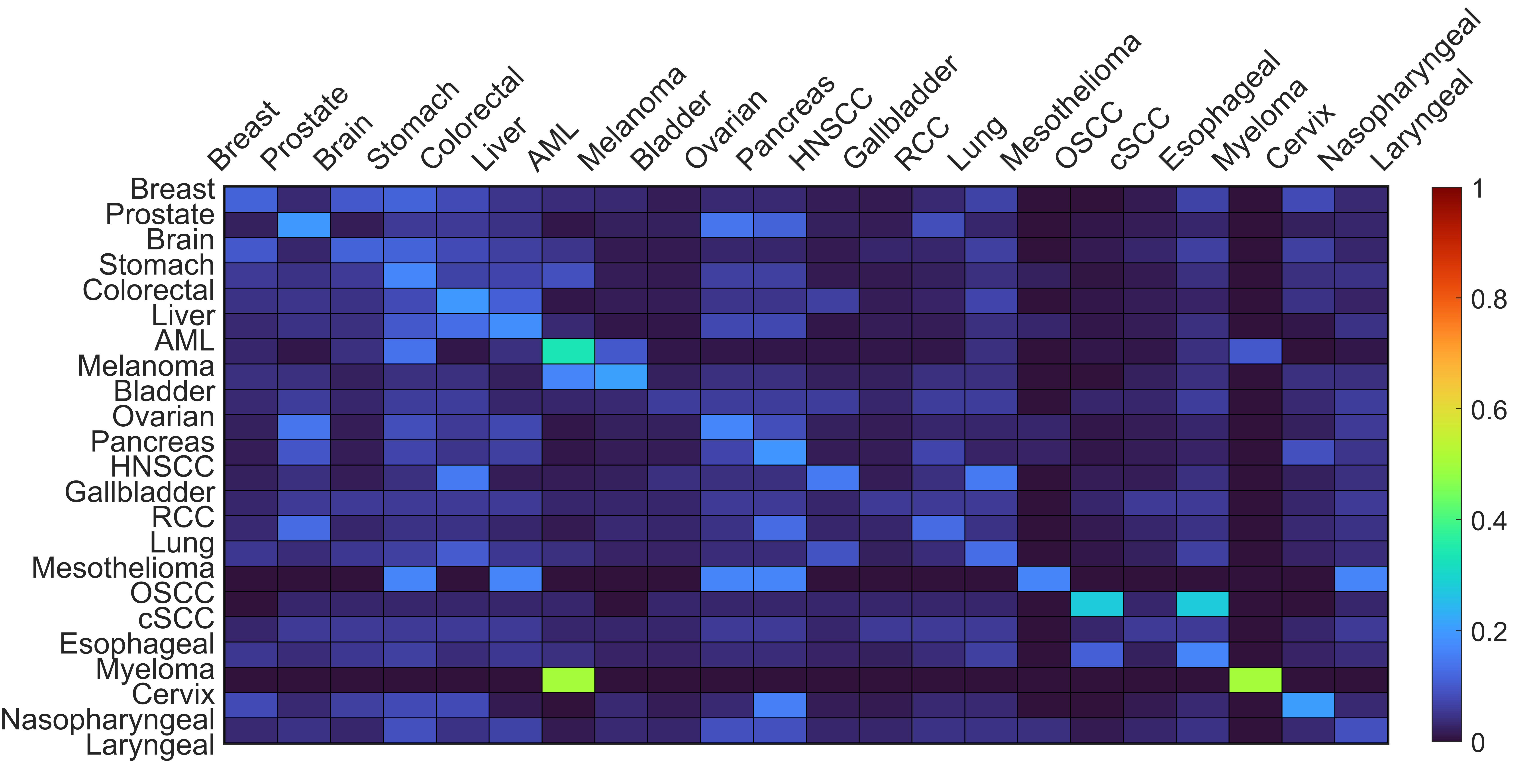}
\caption{\label{fig:pvv} Colour representation of the transition matrix $ P_{uv} $ between organs for the hypergraph $ H $.}
\end{center}
\end{figure}

Fig. (\ref{fig:pvv}) illustrates the transition probabilities between organs, providing insight into potential cancer spread pathways. The matrix highlights key patterns in metastasis, such as the dominance of diagonal elements, which indicates that metastasis often begins in nearby tissue within the same organ \cite{epcam2023}. These findings are supported by studies showing that $EpCAM+$ and $CD90+$ markers are associated with increased risks of metastasis and treatment resistance in hepatocellular carcinoma \cite{epcam2023}. For numerical details of the matrix, see Appendix \ref{Pvvmatrix}. The figure serves as an illustration, as the matrix has large dimensions ($23 \times 23$).

In this analysis, organs such as AML and myeloma show the highest probabilities of internal transitions, whereas organs such as the stomach, colorectal, and liver exhibit elevated probabilities of becoming metastatic sites. These findings suggest that markers such as $CD49+$, $CD90+$, $CD133+$, and $EpCAM+$ are strongly associated with promoting metastasis in these organs \cite{aacr}. Additionally, other organs like prostate, Ovarian, Pancreas, and lung are moderately likely to harbour new cancers, with markers such as $ALDH+$ serving as indicators of metastatic potential \cite{aldh2023}.

We extend this analysis by constructing the hyperedge-to-hyperedge transition matrix (CSCM--CSCM) $P_{ce}$, which models transitions between CSC populations characterised by different markers:

\begin{equation}
P_{ce} = P_{ev} P_{ve}, \label{pee}
\end{equation}

\begin{figure}
\begin{center}
\includegraphics[width=0.95\linewidth]{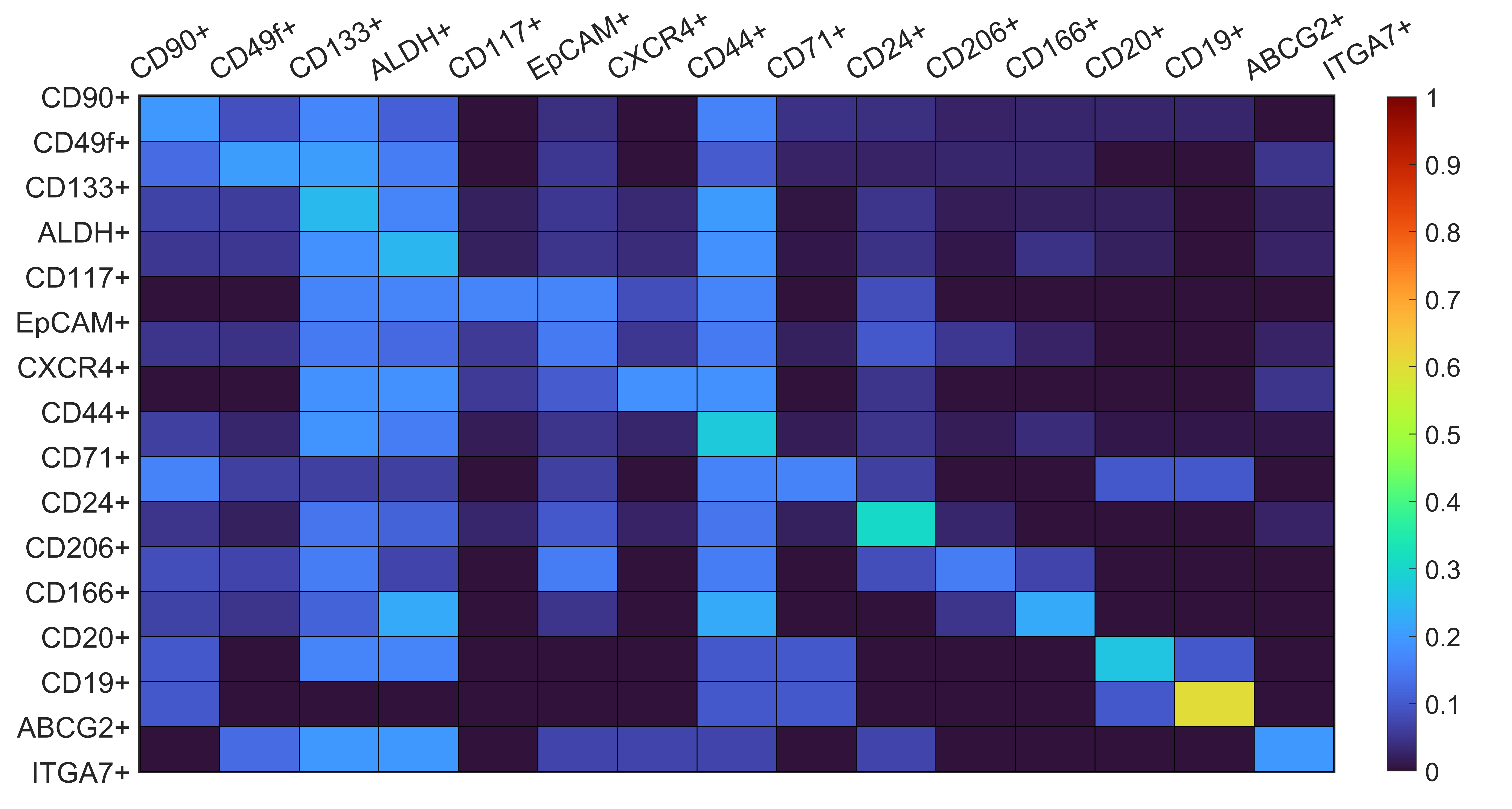}
\caption{\label{fig:pee} Colour representation of the transition matrix $P_{ce}$ between CSCMs for hypergraph $H$.}
\end{center}
\end{figure}

As shown in Fig. (\ref{fig:pee}), this matrix captures the evolution of CSC populations across different phenotypic states defined by markers such as $CD90+$, $CD49f+$, $CD133+$, and $ALDH+$ \cite{Roerink2018}. High-probability transitions between markers, such as from $CD44+$ to $CD133+$, highlight critical pathways in cancer progression. These suggest that cancer cells may adapt their phenotypic traits, acquiring invasive and metastatic capabilities, especially in advanced stages \cite{Batlle2017, Rycaj2015}. For detailed numerical values of the $P_{ce}$ matrix, see Appendix \ref{Peematrix}. Fig. (\ref{fig:pee}) is included for illustrative purposes, as $P_{ce}$ has large dimensions ($16 \times 16$).

The biological relevance of transition matrices lies in their ability to provide new insights into metastasis pathways based on CSCs characterised by CSCMs. By analysing organ-organ and CSCM-CSCM transitions, we can identify the organs most vulnerable to cancer spread and the markers most influential in the process. For example, the high transition probabilities observed in organs like the liver and colorectal tract align with clinical observations of frequent metastasis in these sites, particularly driven by markers such as $CD44+$ and $EpCAM+$ \cite{zarke, kalan}.

Similarly, the CSCM-CSCM matrix $P_{ce}$ reveals key phenotypic shifts in CSC populations, highlighting pathways through which cancer cells may evolve to gain metastatic advantages. This analysis provides critical information for developing strategies to target CSCMs effectively, potentially disrupting metastasis at its source.

\subsection{Stationary distributions}

A stationary distribution in a Markov chain is a probability distribution that remains unchanged over time. If we start with this distribution and repeatedly apply the transition matrix $ P $ (e.g., $ P_{uv} $ or $ P_{ce} $), the probabilities of the states will eventually converge to a steady state  \cite{markov}. This allows us to analyse which organs or CSCMs are most likely to be involved in metastasis over time.

For the transition matrix $ P_{uv} $, which models the interactions between organs on the basis of shared CSCMs, the stationary distribution $\pi_v$ (for vertices) satisfies:

\begin{equation}
\pi_v P_{uv} = \pi_v
\label{estapvv}
\end{equation}

This equation reflects the long-term probabilities that certain organs will accumulate CSCMs.

Similarly, the stationary distribution for CSCMs is given by the matrix $ P_{ce} $, which models the interactions between different markers:

\begin{equation}
\pi_e P_{ce} = \pi_e
\label{estapee}
\end{equation}

Table (\ref{tab:combined}) summarises the stationary distributions for organs ($\pi_v$) and CSCMs ($\pi_e$), providing valuable insights into their involvement in metastasis. The stationary distribution highlights key organs such as the $stomach$, $colorectal$ and $Pancreas$, which have relatively high probabilities of accumulating CSCMs. These findings align with clinical observations of frequent metastasis in these organs  \cite{organ_prob}.

\begin{table}[h]
 \caption{Stationary distributions $ \pi_v $ for organs and $ \pi_e $ for CSCMs.}  
 \label{tab:combined}
 \begin{tabular}{@{}llll@{}}
 \toprule
 \textbf{\textit{Organ}} & \textbf{$\pi_v$} & \textbf{\textit{CSCM}} & \textbf{$\pi_e$} \\ 
 \midrule
 \textit{Breast}         & 0.0426 & $CD90+$         & 0.0745 \\ 
 \textit{Prostate}       & 0.0638 & $CD49f+$        & 0.0532 \\ 
 \textit{Brain}          & 0.0426 & $CD133+$        & 0.1809 \\ 
 \textit{Stomach}        & 0.0851 & $ALDH+$         & 0.1596 \\ 
 \textit{Colorectal}     & 0.0745 & $CD117+$        & 0.0213 \\ 
 \textit{Liver}          & 0.0638 & $EpCAM+$        & 0.0638 \\ 
 \textit{AML}            & 0.0532 & $CXCR4+$        & 0.0319 \\ 
 \textit{Melanoma}       & 0.0319 & $CD44+$         & 0.1915 \\ 
 \textit{Bladder}        & 0.0213 & $CD71+$         & 0.0213 \\ 
 \textit{Ovarian}        & 0.0638 & $CD24+$         & 0.0638 \\ 
 \textit{Pancreas}       & 0.0745 & $CD206+$        & 0.0213 \\ 
 \textit{HNSCC}          & 0.0319 & $CD166+$        & 0.0213 \\ 
 \textit{Gallbladder}     & 0.0426 & $CD20+$         & 0.0213 \\ 
 \textit{RCC}            & 0.0426 & $CD19+$         & 0.0213 \\ 
 \textit{Lung}           & 0.0532 & $ABCG2+$        & 0.0213 \\ 
 \textit{Mesothelioma}   & 0.0106 & $ITGA7+$        & 0.0213 \\ 
 \textit{OSCC}           & 0.0213 & $CD90+$         & 0.0213 \\ 
 \textit{cSCC}           & 0.0213 & $ABCG2+$        & 0.0213 \\ 
 \textit{Esophageal}     & 0.0213 & $CD206+$        & 0.0213 \\ 
 \textit{Myeloma}        & 0.0532 & $ABCG2+$        & 0.0213 \\ 
 \textit{Cervix}         & 0.0106 & $CD44+$         & 0.1915 \\ 
 \textit{Nasopharyngeal} & 0.0213 & $CD20+$         & 0.0213 \\ 
 \bottomrule
 \end{tabular}
\end{table}

Markers such as $CD44+$, $CD133+$ and $ALDH+$, which have high stationary probabilities, are crucial in driving metastasis across multiple organs  \cite{cscm_roles}. Understanding these distributions offers deeper insight into how CSCMs influence cancer spread and highlights potential targets for therapeutic interventions.

\begin{figure}
\begin{center}
\includegraphics[width=1\linewidth]{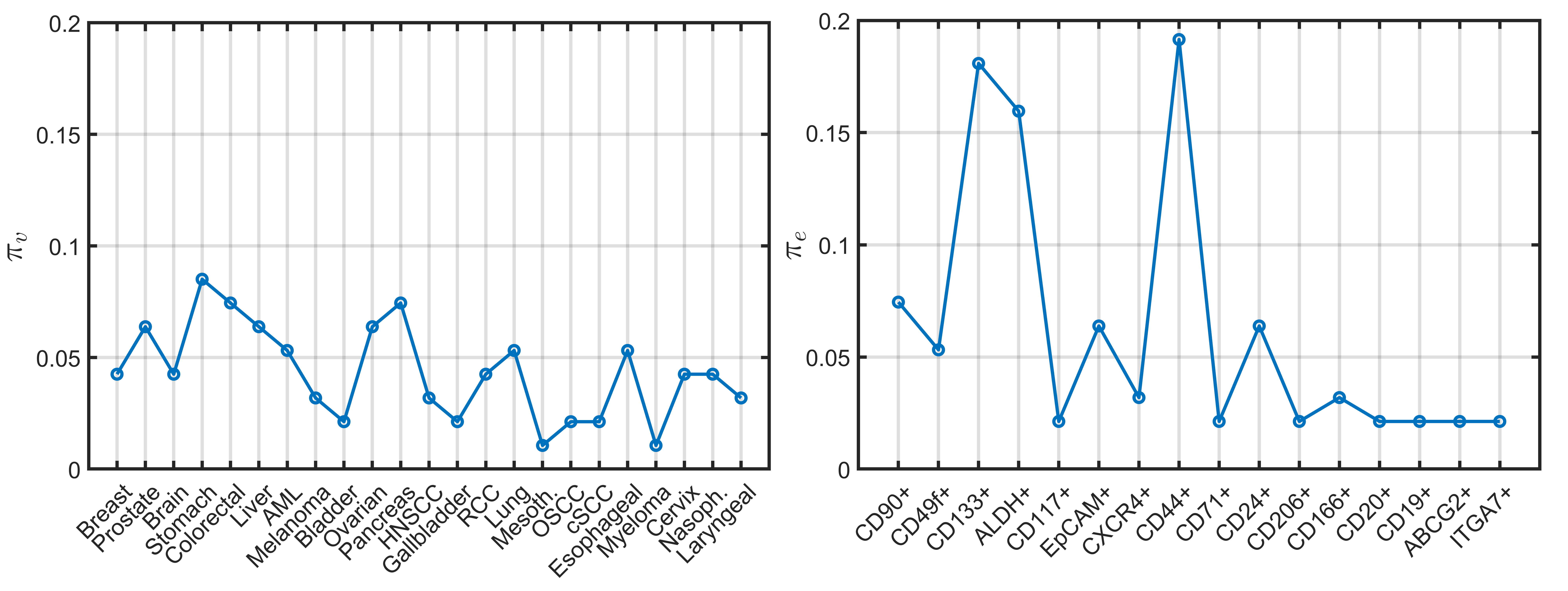}
\caption{\label{fig:estdoble} Representation of the stationary distributions $ \pi_v $ (left) and $ \pi_e $ (right) of the transition matrices $ P_{uv} $ and $ P_{ce} $, respectively.}
\end{center}
\end{figure} 

Finally, constructing transition matrices for higher powers of $P$ (e.g., $ P_{uv}^2 $, $ P_{ce}^2 $) in this context is unnecessary. This is because a body cannot sustain a multi-metastatic process without treatment. For further insights into the branching processes in metastasis, see  \cite{margarit2016} and  \cite{margarit2023}.

\section{Mutual information \label{mutualinfo}}
In information theory, mutual information ($I$) is a fundamental measure of the dependence between two random variables. In the context of hypergraphs -structures that connect multiple nodes- $I$ is crucial for quantifying the relationships between nodes and hyperedges. It serves as a key tool for analysing complex network structures such as hypergraphs \cite{mi}. Put simply, $I$ quantifies the reduction in uncertainty about one variable when another is known. In hypergraphs, it measures the strength of association between nodes and hyperedges, revealing underlying structural or functional patterns.

In our specific hypergraph defined by the incidence matrix $ H $, with nodes represented by $ V $ and hyperedges by $ E $, the mutual information between nodes and hyperedges quantifies their dependence in relation to the presence of specific nodes within particular hyperedges  \cite{sha, mi2}. This process involves calculating the entropies of the nodes $ \mathcal{S}(V) $, of the hyperedges $ \mathcal{S}(E) $ and their joint entropy $ \mathcal{S}(V, E) $  \cite{sha, mi2}. By conducting these calculations, we can determine the mutual information $ I(V; E) $, which facilitates the classification of nodes on the basis of their influence within the hypergraph structure. The calculation involves assessing the marginal probabilities, which represent the likelihood of a specific event occurring independently, for nodes and hyperedges, as well as their joint probabilities.

Shannon entropy ($\mathcal{S}$) for a random variable $X$ with probability distribution $p(x)$ is defined as follows:
$$ \mathcal{S}(X) = -\sum_{x} p(x) \log_2 p(x) $$
This formula calculates the average level of uncertainty in variable $X$. Maximum entropy occurs when all probabilities are equal. For two random variables $X$ and $Y$, their joint entropy $\mathcal{S}(X, Y)$ and mutual information $I(X; Y)$ are defined as:
 $$ \mathcal{S}(X, Y) = -\sum_{x, y} p(x, y) \log_2 p(x, y) \qquad  \mbox{and}\qquad  I(X; Y) = \mathcal{S}(X) + \mathcal{S}(Y) - \mathcal{S}(X, Y) $$ 
These measures quantify the amount of information shared between variables. When applied to our hypergraph, they allow for the calculation of mutual information between nodes and hyperedges. For our analysis of hypergraph $ H $, we compute both marginal and joint probabilities to determine the mutual information. Marginal probabilities:
$$p_v(v) = \frac{\sum_{e} H(v, e)}{\sum_{v,e} H(v, e)} \qquad  \mbox{and}\qquad   p_e(e) = \frac{\sum_{v} H(v, e)}{\sum_{v,e} H(v, e)}$$
$p_v(v)$ is computed as the sum of incidences of node $v$ across all hyperedges, divided by the total sum of incidences across all nodes and hyperedges. Similarly, $p_e(e)$ is calculated as the sum of incidences of hyperedges $e$ across all nodes, divided by the total sum of incidences across all nodes and hyperedges. Joint probabilities:
$$p_{ve}(v, e) = \frac{H(v, e)}{\sum_{v,e} H(v, e)}$$

Here, $p_{ve}(v, e)$ represents the likelihood of a node $v$ and a hyperedge $e$ coexisting in a hypergraph. It is calculated as the specific incidence of node $v$ in hyperedge $e$ divided by the total sum of incidences across all nodes and hyperedges. Thus, mutual information for nodes ($I(v)$) and hyperedges ($I(e)$) is computed as follows:

\begin{equation}
I(v) = \sum_{e} p_{ve}(v, e) \log_2 \left( \frac{p_{ve}(v, e)}{p_v(v) p_e(e)} \right)
\label{Iv}
\end{equation}

\begin{equation}
I(e) = \sum_{v} p_{ve}(v, e) \log_2 \left( \frac{p_{ve}(v, e)}{p_v(v) p_e(e)} \right)
\label{Ie}
\end{equation}
Eqs. (\ref{Iv}) and (\ref{Ie}) are similar, indicating that they represent closely related expressions for both $ I(v) $ and $ I(e) $. This similarity could lead to misinterpretation if the specific context is not taken into account. In this study, let us remember that each node represents an affected organ in cancer, whereas each hyperedge represents a cancer CSCM. The expression for $ I(v) $ calculates the mutual information averaged over all hyperedges involving a specific organ $ v $, whereas the expression for $ I(e) $ calculates the mutual information averaged over all organs involved in a specific CSCM $ e $. The term $ \log_2 \left( \frac{p_{ve}(v, e)}{p_v(v) p_e(e)} \right) $ quantifies the deviation in the probability of a node $ v $ being connected to a hyperedge $ e $ compared with the scenario where $ v $ and $ e $ are independent.

Tables (\ref{tab:clasnodo}) and (\ref{tab:clasar}) summarise the mutual information values for each node and hyperedge, highlighting those with significant associations. To classify nodes and hyperedges more precisely, we divide them into tertiles -groups representing three equal portions of the data- based on the calculated mutual information $I$. Threshold values, $I_{inf}$ and $I_{sup}$, define these tertiles for each variable. Nodes and hyperedges with low mutual information are classified as $I < I_{inf}$, those with medium mutual information as $I_{inf} \leq I \leq I_{sup}$, and those with high mutual information as $I > I_{sup}$. Threshold values are derived from the data range and are as follows: for nodes, $I_{inf} = 0.0494$ and $I_{sup} = 0.0623$; for hyperedges, $I_{inf} = 0.0681$ and $I_{sup} = 0.0836$.

\begin{table}[h]
    \caption{Classification of nodes/organs according to mutual information.}\label{tab:clasnodo}
\begin{tabular}{@{}llllll@{}}
 \toprule        
        \textbf{High} & \textbf{$I$} & \textbf{Medium} & \textbf{$I$} & \textbf{Low} & \textbf{$I$} \\
        \midrule
        \textit{AML} & 0.1190 & \textit{Myeloma} & 0.0591 & \textit{cSCC} & 0.0303 \\
        \textit{Cervix} & 0.0734 & \textit{Esophageal} & 0.0552 & \textit{Nasopharyngeal} & 0.0369 \\
        \textit{Liver} & 0.0700 & \textit{HNSCC} & 0.0558 & \textit{Gallbladder} & 0.0303 \\
        \textit{Colorectal} & 0.0699 & Ovarian & 0.0583 & \textit{Bladder} & 0.0323 \\
        \textit{Prostate} & 0.0690 & \textit{Melanoma} & 0.0629 & \textit{Laryngeal} & 0.0292 \\
         &  & \textit{OSCC} & 0.0632 &  &  \\
         &  & \textit{Lung} & 0.0490 &  &  \\
         &  & \textit{RCC} & 0.0475 &  &  \\
         &  & \textit{Mesothelioma} & 0.0422 &  &  \\
         &  & \textit{Stomach} & 0.0529 &  &  \\
         &  & \textit{Brain} & 0.0514 &  &  \\
         &  & \textit{Breast} & 0.0542 &  &  \\
                 \hline
    \end{tabular}
\end{table}

\begin{table}[h]  
    \caption{Classification of hyperedges/CSCMs according to mutual information.}\label{tab:clasar}
\begin{tabular}{@{}llllll@{}} 
 \toprule               
        \textbf{High} & \textbf{$I$} & \textbf{Medium} & \textbf{$I$} & \textbf{Low} & \textbf{$I$} \\
           \midrule
       \textit{ CD24+} & 0.1153 & \textit{EpCAM+ }& 0.0792 &\textit{ CD133+} & 0.0613 \\
       \textit{ CD90+} & 0.1030 & \textit{CXCR4+} & 0.0800 & \textit{CD117+} & 0.0632 \\
       \textit{ CD49f+} & 0.0995 & \textit{ALDH+} & 0.0801 &\textit{ ABCG2+} & 0.0670 \\
        \textit{CD19+ }& 0.0935 & \textit{CD20+} & 0.0766 & \textit{CD71+} & 0.0616 \\
      \textit{  CD166+} & 0.0872 & \textit{CD44+} & 0.0683 & \textit{CD206+} & 0.0608 \\
      \textit{  ITGA7+} & 0.0828 &  &  &  &  \\
        \hline
    \end{tabular}
\end{table}

The analysis of mutual information ($I$) reveals significant associations between CSCMs and various cancer types. Markers such as $CD24+$, $CD90+$ and $CD49f+$ present high $I$ values within the CSCM hypergraphs, indicating a strong link to tumorigenesis. Recent studies emphasise the critical role of these markers in regulating key signalling pathways, such as the Wnt and Notch pathways, which are essential for self-renewal and treatment resistance in multiple cancers \cite{Visvader2008, Kwon2020, Chen2013, Le2020}.

Node classification on the basis of $I$ values shows that organs such as the $AML$, $cervix$ and $liver$ are highly dependent on markers such as $CD24+$ and $CD90+$. These findings identify these CSC sub-populations as pivotal in cancer initiation and prevention within specific contexts \cite{Kwon2020, Chen2013}. Conversely, the lower association in organs such as $cSCC$ and $nasopharyngeal$ tissue points to a reduced relevance of these markers in the pathogenesis of these cancers, as corroborated by transcriptomic and differential gene expression analyses \cite{Le2020}. Identifying high-$I$ markers such as $CD24+$ and $CD90+$ offers key insights into tumour cell characteristics and suggests new avenues for targeted therapies. Such therapies, combined with early biomarker identification, could improve treatment outcomes and mitigate the drug resistance often observed in conventional approaches \cite{OBrien2012}.


\section{Conclusions\label{conclusiones}}

This study integrates hypergraph-based models with computational tools, including Markov chains and mutual information analyses, to provide a comprehensive framework for understanding CSCM dynamics and their role in cancer progression. Our approach uncovers critical metastatic pathways and highlights the interdependencies between CSCMs and organs, offering novel insights into the mechanisms driving tumour heterogeneity and metastasis.

Key findings include variations in organ susceptibility, with $breast$, $stomach$, and $colorectal$ regions exhibiting higher probabilities in the stationary distribution $\pi_v$, reflecting organ-specific factors that influence cancer dynamics \cite{smith2023, johnson2022}. Similarly, the stationary distribution $\pi_e$ for CSCMs identifies pivotal markers like $CD44+$ and $CD133+$ as central to cancer initiation and progression \cite{taylor2022, brown2023}. Transition matrices further reveal phenotypic transitions and metastatic pathways mediated by key CSCMs, shedding light on their critical roles in cancer invasion \cite{clark2022, harris2023}.

Mutual information analysis complements these findings by uncovering significant dependencies between organs and CSCMs, reinforcing the concept of co-regulation and interconnectivity in cancer dynamics \cite{manni2022}. Persistent markers such as $ALDH+$ emerge as crucial components in metastasis, aligning with studies that emphasise the value of information theory for characterising complex biological networks \cite{gomez2023, rodriguez2022}.

Despite promising results, certain limitations remain. The static nature of the data constrains our ability to fully capture CSCM temporal dynamics. However, our computational framework lays the groundwork for addressing these challenges through future integration of dynamic datasets and advanced machine learning techniques.


This study highlights the potential of hypergraph-based models to capture the complexity of CSCM distribution across multiple organs. By integrating Markov chains and mutual information analyses, we outline a probabilistic framework that infers potential metastatic pathways and uncovers patterns of CSCM interdependence. These findings provide a foundational perspective for future studies aiming to characterise CSCM dynamics and explore targeted interventions to disrupt metastasis at its source. Our key findings include the following:

The stationary distribution $\pi_v$ highlights variations in cancer susceptibility across organs, with the $breast$, $stomach$, and $colorectal$ regions showing higher probabilities. These findings reflect the complex interplay between organ-specific factors and cancer progression \cite{smith2023, johnson2022}. The transition matrix $P_{uv}$ uncovers potential metastasis pathways, enhancing our understanding of how cancer spreads (by circulating CSCs) between organs \cite{anderson2022}. Similarly, the stationary distribution $\pi_e$ for CSCMs identifies the critical roles of markers such as $CD44+$ and $CD133+$ in cancer initiation and progression across multiple organs \cite{taylor2022, brown2023}.

Analysis of the transition matrix $P_{ce}$ reveals transitions between phenotypic states, suggesting important pathways for cancer invasion, including those linked to $CD44+$ and $CD133+$ markers \cite{clark2022, harris2023}. Our stochastic modelling approach simulates CSCM interactions dynamically, providing insights into the underlying mechanisms of their spread \cite{lee2022}. Additionally, mutual information analysis uncovers significant dependencies between organs and CSCMs, reinforcing the concept of co-regulation in cancer dynamics \cite{manni2022}.

Markers such as $CD44+$ and $CD133+$ are central to tumour heterogeneity, supporting studies that highlight the value of mutual information in characterising complex biological networks \cite{gomez2023, rodriguez2022}. The distribution patterns of CSCMs across organs provide insights into metastasis and cancer progression, particularly regarding the persistence of these markers, including $ALDH+$.

While the findings are promising, certain limitations should be noted. For instance, the quality and availability of CSCM data may impact the accuracy of our hypergraph models. Moreover, the static nature of the data may not fully capture the temporal dynamics of CSCMs, although the use of transition matrices has allowed us to gain valuable insights into these processes.

Looking forward, future research should explore the integration of hypergraph theory with advanced machine learning techniques to improve predictive models of cancer progression and enable more accurate identification of metastatic pathways. Emphasising structural features such as higher-order connectivity and organ-specific marker distributions could offer new insights into the organisation and dynamics of CSCM interactions, ultimately contributing to more effective therapeutic strategies.

\section*{Acknowledgements}

This work is financed by PIP N$^o$ 11220200100439CO of the Consejo Nacional de Investigaciones Cient\'ificas y T\'ecnicas (CONICET), Argentina. 

Funded by the European Union, views and opinions expressed are however those of the author(s) only and do not necessarily reflect those of the European Union or European Research Executive Agency (REA) (granting authority). Neither the European Union nor the granting authority can be held responsible for them. The project leading to this application has received funding from the European Union's Horizon Europe programme under the MSCA-SE grant agreement N$^o$ 101131463 - SIMBAD: Statistical Inference from Multiscale Biological Data: Theory, Algorithms, Applications.


%
%
%

\appendix 
\begin{landscape}
\section{Annexes \label{matrices}}

\subsection{Incidence matrix $H$\label{Hmatrix}}
\begin{spacing}{.75}\setlength{\arraycolsep}{0.05cm}
\begin{small}

\begin{equation}
H\!=\!\kbordermatrix{
        ~ & CD90+ & CD49f+ & CD133+ & ALDH+ & CD117+ & EpCAM+ & CXCR4+ & CD44+ & CD71+ & CD24+ & CD206+ & CD166+ & CD20+ & CD19+ & ABCG2+ & ITGA7+ \cr \smallskip
        Breast & 1 & 1 & 1 & 1 & 0 & 0 & 0 & 0 & 0 & 0 & 0 & 0 & 0 & 0 & 0 & 0\cr \smallskip
        Prostate & 0 & 0 & 1 & 1 & 1 & 1 & 1 & 1 & 0 & 0 & 0 & 0 & 0 & 0 & 0 & 0\cr \smallskip
        Brain & 1 & 1 & 1 & 0 & 0 & 0 & 0 & 1 & 0 & 0 & 0 & 0 & 0 & 0 & 0 & 0\cr \smallskip
        Stomach & 1 & 1 & 1 & 1 & 0 & 1 & 0 & 1 & 1 & 1 & 0 & 0 & 0 & 0 & 0& 0\cr \smallskip
        Colorectal & 0 & 1 & 1 & 1 & 0 & 1 & 0 & 1 & 0 & 0 & 1 & 1 & 0 & 0 & 0& 0\cr \smallskip
        Liver & 1 & 0 & 1 & 0 & 0 & 1 & 0 & 1 & 0 & 1 & 1 & 0 & 0 & 0 & 0& 0\cr \smallskip
        AML & 1 & 0 & 0 & 0 & 0 & 0 & 0 & 1 & 1 & 0 & 0 & 0 & 1 & 1 & 0& 0\cr \smallskip
        Melanoma & 0 & 0 & 1 & 1 & 0 & 0 & 0 & 0 & 0 & 0 & 0 & 0 & 1 & 0 & 0& 0\cr \smallskip
        Bladder & 0 & 0 & 0 & 1 & 0 & 0 & 0 & 1 & 0 & 0 & 0 & 0 & 0 & 0 & 0& 0\cr \smallskip
        Ovarian & 0 & 0 & 1 & 1 & 1 & 1 & 0 & 1 & 0 & 1 & 0 & 0 & 0 & 0 & 0& 0\cr \smallskip
        Pancreas & 0 & 0 & 1 & 1 & 0 & 1 & 1 & 1 & 0 & 1 & 0 & 0 & 0 & 0 & 1& 0\cr \smallskip
        HNSCC & 0 & 0 & 0 & 1 & 0 & 0 & 0 & 1 & 0 & 0 & 0 & 1 & 0 & 0 & 0& 0\cr \smallskip
        Gallbladder & 0 & 0 & 1 & 0 & 0 & 0 & 0 & 1 & 0 & 0 & 0 & 0 & 0 & 0 & 0& 0\cr \smallskip
        RCC & 0 & 0 & 1 & 1 & 0 & 0 & 1 & 1 & 0 & 0 & 0 & 0 & 0 & 0 & 0& 0\cr \smallskip
        Lung & 1 & 0 & 1 & 1 & 0 & 0 & 0 & 1 & 0 & 0 & 0 & 1 & 0 & 0 & 0& 0\cr \smallskip
        Mesothelioma & 0 & 0 & 0 & 0 & 0 & 0 & 0 & 0 & 0 & 1 & 0 & 0 & 0 & 0 & 0& 0\cr \smallskip
        OSCC & 0 & 0 & 0 & 0 & 0 & 0 & 0 & 1 & 0 & 0 & 0 & 0 & 0 & 0 & 0& 1\cr \smallskip
        cSCC & 0 & 0 & 1 & 0 & 0 & 0 & 0 & 1 & 0 & 0 & 0 & 0 & 0 & 0 & 0& 0\cr \smallskip
        Esophageal & 1 & 0 & 1 & 1 & 0 & 0 & 0 & 1 & 0 & 0 & 0 & 0 & 0 & 0 & 0& 1\cr \smallskip
        Myeloma & 0 & 0 & 0 & 0 & 0 & 0 & 0 & 0 & 0 & 0 & 0 & 0 & 0 & 1 & 0& 0\cr \smallskip
        Cervix & 0 & 1 & 1 & 1 & 0 & 0 & 0 & 0 & 0 & 0 & 0 & 0 & 0 & 0 & 1& 0\cr \smallskip
        Nasopharyngeal & 0 & 0 & 1 & 1 & 0 & 0 & 0 & 1 & 0 & 1 & 0 & 0 & 0 & 0 & 0& 0\cr 
        Laryngeal & 0 & 0 & 1 & 1 & 0 & 0 & 0 & 1 & 0 & 0 & 0 & 0 & 0 & 0 & 0& 0 }
\label{Hext}
\end{equation}  
\end{small}
\end{spacing}\end{landscape}

\subsection{Transition Matrix $P_{vv}$\label{Pvvmatrix}}
\begin{spacing}{1.25}\setlength{\arraycolsep}{0.02cm}

\begin{small}

\begin{equation}
P_{uv}=\kbordermatrix{
  &  v_1 & v_2 &  v_3 &  v_4 &  v_5 &  v_6 &  v_7 &  v_8 &  v_9 &  v_{10} & v_{11} & v_{12} & v_{13} & v_{14} &  v_{15} & v_{16} & v_{17} &  v_{18} &  v_{19} &  v_{20} & v_{21} & v_{22} &  v_{23} \\
   v_1 & \frac{29}{248} & \frac{8}{255} & \frac{24}{239} & \frac{29}{248} & \frac{21}{258} & \frac{12}{239} & \frac{5}{140} & \frac{8}{255} & \frac{1}{60} & \frac{8}{255} & \frac{8}{255} & \frac{1}{60} & \frac{1}{68} & \frac{8}{255} & \frac{17}{253} & 0 & 0 & \frac{1}{68} & \frac{17}{253} & 0 & \frac{21}{258} & \frac{8}{255} & \frac{8}{255} \\
   v_2 & \frac{1}{48} & \frac{49}{249} & \frac{1}{52} & \frac{4}{69} & \frac{4}{69} & \frac{29}{619} & \frac{1}{108} & \frac{1}{48} & \frac{11}{540} & \frac{13}{92} & \frac{13}{113} & \frac{11}{540} & \frac{1}{52} & \frac{43}{501} & \frac{1}{33} & 0 & \frac{1}{108} & \frac{1}{52} & \frac{1}{33} & 0 & \frac{1}{48} & \frac{1}{33} & \frac{1}{33} \\
   v_3 & \frac{24}{239} & \frac{2}{70} & \frac{7}{61} & \frac{7}{61} & \frac{19}{242} & \frac{7}{109} & \frac{4}{81} & \frac{1}{68} & \frac{1}{72} & \frac{2}{70} & \frac{2}{70} & \frac{1}{72} & \frac{2}{70} & \frac{2}{70} & \frac{7}{109} & 0 & \frac{1}{72} & \frac{2}{70} & \frac{7}{109} & 0 & \frac{11}{170} & \frac{2}{70} & \frac{2}{70} \\
   v_4 & \frac{3}{51} & \frac{2}{46} & \frac{7}{122} & \frac{12}{71} & \frac{3}{45} & \frac{12}{163} & \frac{4}{46} & \frac{1}{64} & \frac{5}{327} & \frac{4}{59} & \frac{4}{59} & \frac{5}{327} & \frac{1}{70} & \frac{3}{91} & \frac{4}{99} & \frac{5}{240} & \frac{1}{144} & \frac{1}{70} & \frac{4}{99} & 0 & \frac{1}{56} & \frac{2}{46} & \frac{3}{91} \\
   v_5 & \frac{7}{151} & \frac{3}{46} & \frac{4}{89} & \frac{4}{51} & \frac{17}{86} & \frac{13}{117} & \frac{1}{126} & \frac{7}{391} & \frac{13}{744} & \frac{3}{46} & \frac{3}{46} & \frac{3}{46} & \frac{1}{61} & \frac{7}{271} & \frac{7}{95} & 0 & \frac{1}{126} & \frac{1}{61} & \frac{7}{271} & 0 & \frac{7}{151} & \frac{7}{271} & \frac{7}{271} \\
   v_6 & \frac{8}{238} & \frac{7}{229} & \frac{4}{85} & \frac{4}{83} & \frac{21}{139} & \frac{3}{73} & \frac{7}{225} & \frac{1}{102} & \frac{8}{855} & \frac{7}{94} & \frac{7}{94} & \frac{7}{94} & \frac{1}{52} & \frac{1}{52} & \frac{1}{23} & \frac{5}{180} & \frac{1}{108} & \frac{1}{52} & \frac{1}{23} & 0 & \frac{1}{102} & \frac{7}{229} & \frac{1}{52} \\
  v_7 & \frac{1}{35} & \frac{1}{90} & \frac{5}{126} & \frac{19}{136} & \frac{1}{90} & \frac{5}{126} & \frac{1}{3} & \frac{1}{10} & \frac{1}{90} & \frac{1}{90} & \frac{1}{90} & \frac{1}{90} & \frac{1}{90} & \frac{1}{90} & \frac{5}{126} & 0 & \frac{1}{90} & \frac{1}{90} & \frac{5}{126} & \frac{1}{10} & 0 & \frac{1}{90} & \frac{1}{90} \\
   v_8 & \frac{2}{49} & \frac{2}{49} & \frac{1}{51} & \frac{2}{49} & \frac{2}{49} & \frac{1}{51} & \frac{5}{30} & \frac{12}{65} & \frac{2}{90} & \frac{2}{49} & \frac{2}{49} & \frac{2}{90} & \frac{1}{51} & \frac{2}{49} & \frac{2}{49} & 0 & 0 & \frac{1}{51} & \frac{2}{49} & 0 & \frac{2}{49} & \frac{2}{49} & \frac{2}{49} \\
   v_9 & \frac{1}{30} & \frac{3}{49} & \frac{1}{36} & \frac{3}{49} & \frac{3}{49} & \frac{1}{36} & \frac{1}{36} & \frac{1}{30} & \frac{3}{49} & \frac{3}{49} & \frac{3}{49} & \frac{3}{49} & \frac{1}{36} & \frac{1}{36} & \frac{1}{36} & \frac{1}{30} & 0 & \frac{1}{36} & \frac{1}{36} & 0 & \frac{1}{30} & \frac{1}{36} & \frac{1}{36} \\
   v_{10} & \frac{2}{47} & \frac{1}{54} & \frac{1}{54} & \frac{2}{47} & \frac{2}{47} & \frac{1}{54} & \frac{1}{54} & \frac{2}{47} & \frac{1}{54} & \frac{1}{54} & \frac{1}{54} & \frac{1}{54} & \frac{1}{54} & \frac{2}{47} & \frac{1}{54} & 0 & 0 & \frac{1}{54} & \frac{2}{47} & 0 & \frac{2}{47} & \frac{2}{47} & \frac{2}{47} \\
   v_{11} & \frac{1}{53} & \frac{1}{53} & \frac{1}{58} & \frac{1}{53} & \frac{1}{53} & \frac{1}{58} & \frac{2}{30} & \frac{1}{33} & \frac{1}{58} & \frac{1}{53} & \frac{1}{53} & \frac{1}{58} & \frac{1}{58} & \frac{1}{53} & \frac{1}{58} & 0 & 0 & \frac{1}{58} & \frac{1}{53} & 0 & \frac{1}{53} & \frac{1}{53} & \frac{1}{53} \\
   v_{12} & \frac{2}{50} & \frac{1}{66} & \frac{1}{65} & \frac{2}{50} & \frac{2}{50} & \frac{1}{65} & \frac{1}{65} & \frac{2}{50} & \frac{1}{65} & \frac{1}{65} & \frac{1}{65} & \frac{1}{65} & \frac{1}{65} & \frac{2}{50} & \frac{1}{65} & 0 & 0 & \frac{1}{65} & \frac{2}{50} & 0 & \frac{2}{50} & \frac{2}{50} & \frac{2}{50} \\
   v_{13} & \frac{1}{67} & \frac{1}{67} & \frac{1}{67} & \frac{1}{67} & \frac{1}{67} & \frac{1}{67} & \frac{1}{67} & \frac{1}{67} & \frac{1}{67} & \frac{1}{67} & \frac{1}{67} & \frac{1}{67} & \frac{1}{67} & \frac{1}{67} & \frac{1}{67} & 0 & 0 & \frac{1}{67} & \frac{1}{67} & 0 & \frac{1}{67} & \frac{1}{67} & \frac{1}{67} \\
   v_{14} & \frac{2}{59} & \frac{1}{75} & \frac{1}{78} & \frac{2}{59} & \frac{2}{59} & \frac{1}{78} & \frac{1}{78} & \frac{2}{59} & \frac{1}{78} & \frac{1}{78} & \frac{1}{78} & \frac{1}{78} & \frac{1}{78} & \frac{2}{59} & \frac{1}{78} & 0 & 0 & \frac{1}{78} & \frac{2}{59} & 0 & \frac{2}{59} & \frac{2}{59} & \frac{2}{59} \\
   v_{15} & \frac{2}{51} & \frac{1}{60} & \frac{1}{61} & \frac{2}{51} & \frac{2}{51} & \frac{1}{61} & \frac{1}{61} & \frac{2}{51} & \frac{1}{61} & \frac{1}{61} & \frac{1}{61} & \frac{1}{61} & \frac{1}{61} & \frac{2}{51} & \frac{1}{61} & 0 & 0 & \frac{1}{61} & \frac{2}{51} & 0 & \frac{2}{51} & \frac{2}{51} & \frac{2}{51} \\
   v_{16} & \frac{2}{55} & \frac{1}{70} & \frac{1}{71} & \frac{2}{55} & \frac{2}{55} & \frac{1}{71} & \frac{1}{71} & \frac{2}{55} & \frac{1}{71} & \frac{1}{71} & \frac{1}{71} & \frac{1}{71} & \frac{1}{71} & \frac{2}{55} & \frac{1}{71} & 0 & 0 & \frac{1}{71} & \frac{2}{55} & 0 & \frac{2}{55} & \frac{2}{55} & \frac{2}{55} \\
   v_{17} & \frac{2}{52} & \frac{1}{65} & \frac{1}{66} & \frac{2}{52} & \frac{2}{52} & \frac{1}{66} & \frac{1}{66} & \frac{2}{52} & \frac{1}{66} & \frac{1}{66} & \frac{1}{66} & \frac{1}{66} & \frac{1}{66} & \frac{2}{52} & \frac{1}{66} & 0 & 0 & \frac{1}{66} & \frac{2}{52} & 0 & \frac{2}{52} & \frac{2}{52} & \frac{2}{52} \\
   v_{18} & \frac{2}{54} & \frac{1}{66} & \frac{1}{67} & \frac{2}{54} & \frac{2}{54} & \frac{1}{67} & \frac{1}{67} & \frac{2}{54} & \frac{1}{67} & \frac{1}{67} & \frac{1}{67} & \frac{1}{67} & \frac{1}{67} & \frac{2}{54} & \frac{1}{67} & 0 & 0 & \frac{1}{67} & \frac{2}{54} & 0 & \frac{2}{54} & \frac{2}{54} & \frac{2}{54} \\
   v_{19} & \frac{2}{53} & \frac{1}{63} & \frac{1}{64} & \frac{2}{53} & \frac{2}{53} & \frac{1}{64} & \frac{1}{64} & \frac{2}{53} & \frac{1}{64} & \frac{1}{64} & \frac{1}{64} & \frac{1}{64} & \frac{1}{64} & \frac{2}{53} & \frac{1}{64} & 0 & 0 & \frac{1}{64} & \frac{2}{53} & 0 & \frac{2}{53} & \frac{2}{53} & \frac{2}{53} \\
   v_{20} & \frac{2}{55} & \frac{1}{67} & \frac{1}{68} & \frac{2}{55} & \frac{2}{55} & \frac{1}{68} & \frac{1}{68} & \frac{2}{55} & \frac{1}{68} & \frac{1}{68} & \frac{1}{68} & \frac{1}{68} & \frac{1}{68} & \frac{2}{55} & \frac{1}{68} & 0 & 0 & \frac{1}{68} & \frac{2}{55} & 0 & \frac{2}{55} & \frac{2}{55} & \frac{2}{55} \\
   v_{21} & \frac{2}{54} & \frac{1}{68} & \frac{1}{69} & \frac{2}{54} & \frac{2}{54} & \frac{1}{69} & \frac{1}{69} & \frac{2}{54} & \frac{1}{69} & \frac{1}{69} & \frac{1}{69} & \frac{1}{69} & \frac{1}{69} & \frac{2}{54} & \frac{1}{69} & 0 & 0 & \frac{1}{69} & \frac{2}{54} & 0 & \frac{2}{54} & \frac{2}{54} & \frac{2}{54} \\
 v_{22} & \frac{2}{53} & \frac{1}{65} & \frac{1}{66} & \frac{2}{53} & \frac{2}{53} & \frac{1}{66} & \frac{1}{66} & \frac{2}{53} & \frac{1}{66} & \frac{1}{66} & \frac{1}{66} & \frac{1}{66} & \frac{1}{66} & \frac{2}{53} & \frac{1}{66} & 0 & 0 & \frac{1}{66} & \frac{2}{53} & 0 & \frac{2}{53} & \frac{2}{53} & \frac{2}{53} \\
 v_{23} & \frac{2}{54} & \frac{1}{67} & \frac{1}{68} & \frac{2}{54} & \frac{2}{54} & \frac{1}{68} & \frac{1}{68} & \frac{2}{54} & \frac{1}{68} & \frac{1}{68} & \frac{1}{68} & \frac{1}{68} & \frac{1}{68} & \frac{2}{54} & \frac{1}{68} & 0 & 0 & \frac{1}{68} & \frac{2}{54} & 0 & \frac{2}{54} & \frac{2}{54} & \frac{2}{54} }
\label{pvvext}
\end{equation}

\end{small}
\end{spacing}

Here, the numbers were rounded to be represented rationally, because they were real and extensive float types. On the other hand, to save space, the names of the organs were expressed based on their vertex numbers.

The vertices are: $V=\{ v_1,\dots,v_{23} \} =\{$\textit{breast, prostate, brain, stomach, colorectal, liver, AML, melanoma, bladder, Ovarian, Pancreas, HNSCC, Gallbladder, RCC, lung, mesothelioma, OSCC, cSCC, esophageal, myeloma, cervix, nasopharyngeal, laryngeal}$\}$.

\subsection{Transition Matrix $P_{ce}$\label{Peematrix}}

\begin{spacing}{1.25}
\begin{small}
\begin{equation}
P_{ce}=\kbordermatrix{
& e_1 & e_2 & e_3 & e_4 & e_5 & e_6 & e_7 & e_8 & e_9 & e_{10} & e_{11} & e_{12} & e_{13} & e_{14} & e_{15} & e_{16} \\
e_1 & \frac{40}{201} & \frac{8}{89} & \frac{34}{200} & \frac{31}{280} & 0 & \frac{1}{24} & 0 & \frac{25}{153} & \frac{13}{280} & \frac{1}{24} & \frac{5}{209} & \frac{2}{71} & \frac{2}{71} & \frac{2}{71} & 0 & \frac{2}{71} \\
e_2 & \frac{1}{8} & \frac{29}{143} & \frac{29}{143} & \frac{11}{72} & 0 & \frac{3}{56} & 0 & \frac{29}{280} & \frac{1}{40} & \frac{1}{40} & \frac{2}{71} & \frac{2}{71} & 0 & 0 & \frac{1}{20} & 0 \\
e_3 & \frac{68}{967} & \frac{37}{617} & \frac{54}{217} & \frac{201}{1214} & \frac{10}{509} & \frac{3}{56} & \frac{17}{515} & \frac{199}{996} & \frac{1}{136} & \frac{71}{1410} & \frac{31}{1704} & \frac{29}{1440} & \frac{10}{509} & 0 & \frac{1}{43} & \frac{2}{170} \\
e_4 & \frac{31}{600} & \frac{11}{216} & \frac{78}{417} & \frac{327}{1344} & \frac{20}{900} & \frac{49}{988} & \frac{29}{780} & \frac{99}{528} & \frac{1}{120} & \frac{46}{1007} & \frac{10}{105} & \frac{68}{1512} & \frac{20}{900} & 0 & \frac{40}{1512} & \frac{13}{960} \\
e_5 & 0 & 0 & \frac{1}{6} & \frac{1}{6} & \frac{1}{6} & \frac{1}{6} & \frac{1}{12} & \frac{1}{6} & 0 & \frac{1}{12} & 0 & 0 & 0 & 0 & 0 & 0 \\
e_6 & \frac{7}{144} & \frac{8}{179} & \frac{68}{449} & \frac{31}{250} & \frac{5}{90} & \frac{68}{449} & \frac{14}{427} & \frac{68}{449} & \frac{5}{240} & \frac{457}{4560} & \frac{14}{271} & \frac{2}{83} & 0 & 0 & \frac{2}{83} & 0 \\
e_7 & 0 & 0 & \frac{56}{300} & \frac{56}{300} & \frac{5}{90} & \frac{23}{210} & \frac{56}{300} & \frac{56}{300} & 0 & \frac{10}{210} & 0 & 0 & 0 & 0 & \frac{10}{210} & 0 \\
e_8 & \frac{28}{441} & \frac{4}{139} & \frac{199}{1056} & \frac{25}{160} & \frac{1}{54} & \frac{17}{336} & \frac{49}{1576} & \frac{142}{519} & \frac{1}{55} & \frac{47}{993} & \frac{43}{2506} & \frac{77}{2048} & \frac{1}{90} & \frac{1}{90} & \frac{5}{632} & \frac{8}{205} \\
e_9 & \frac{13}{80} & \frac{1}{16} & \frac{1}{16} & \frac{1}{16} & 0 & \frac{1}{16} & 0 & \frac{13}{80} & \frac{13}{80} & \frac{1}{16} & 0 & 0 & \frac{1}{10} & \frac{1}{10} & 0 & 0 \\
e_{10} & \frac{7}{144} & \frac{5}{24} & \frac{64}{451} & \frac{41}{359} & \frac{5}{180} & \frac{457}{4560} & \frac{2}{84} & \frac{64}{451} & \frac{5}{240} & \frac{1523}{4950} & \frac{5}{180} & 0 & 0 & 0 & \frac{2}{84} & 0 \\
e_{11} & \frac{10}{120} & \frac{10}{120} & \frac{25}{162} & \frac{10}{120} & 0 & \frac{25}{162} & 0 & \frac{25}{162} & 0 & \frac{10}{120} & \frac{25}{162} & \frac{10}{120} & 0 & 0 & 0 & 0 \\
e_{12} & \frac{10}{150} & \frac{5}{105} & \frac{13}{114} & \frac{49}{218} & 0 & \frac{5}{105} & 0 & \frac{49}{218} & 0 & 0 & \frac{5}{105} & \frac{49}{218} & 0 & 0 & 0 & 0 \\
e_{13} & \frac{15}{150} & 0 & \frac{1}{6} & \frac{1}{6} & 0 & 0 & 0 & \frac{15}{150} & \frac{15}{150} & 0 & 0 & 0 & \frac{4}{15} & \frac{15}{150} & 0 & 0 \\
e_{14} & \frac{15}{150} & 0 & 0 & 0 & 0 & 0 & 0 & \frac{15}{150} & \frac{15}{150} & 0 & 0 & 0 & \frac{15}{150} & \frac{9}{15} & 0 & 0 \\
e_{15} & 0 & \frac{1}{8} & \frac{28}{143} & \frac{28}{143} & 0 & \frac{1}{14} & \frac{1}{14} & \frac{1}{14} & 0 & \frac{1}{14} & 0 & 0 & 0 & 0 & \frac{28}{143} & 0 \\
e_{16} & \frac{15}{150} & 0 & \frac{15}{150} & \frac{15}{150} & 0 & 0 & 0 & 0 & 0 & 0 & 0 & 0 & 0 & 0 & 0 & 0 }
\label{peeext}
\end{equation}  
\end{small}
\end{spacing}

In this case, again, the numbers were rounded to be represented rationally, because they were real and extensive float types. The names of the CSCMs were expressed based on their hyperedges numbers.

The hyperedges are: $E=\{e_1, \dots , e_{16}  \} =\{$\textit{CD90+, CD49f+, CD133+, ALDH+, CD117+, EpCAM+, CXCR4+, CD44+, CD71+, CD24+, CD206+, CD166+, CD20+, CD19+, ABCG2+}$\}$.


\clearpage
\subsection{Algorithm for CSCM Hypergraph Analysis\label{algoritmo}}

The algorithm \ref{algor} summarises the sequence of computational steps used to analyse the CSCM hypergraph and its associated information-theoretic measures. All procedures were implemented in Matlab 2023b \cite{matlab}.

\begin{small}

\begin{algorithm}[H]
\SetAlgoNoLine
\SetInd{0em}{1.5em} 
\caption{Computational steps for CSCM Hypergraph Analysis\label{algor}}\medskip

\KwData{Incidence matrix $H \in \{0,1\}^{23 \times 16}$ (organs $\times$ CSCMs)}
\KwResult{Transition matrices, stationary distributions, and mutual information values}

\vspace{0.5em}
\textbf{\textit{Step 1: Construct hypergraph structure}}

Define sets of organs $V = \{v_1,\dots,v_{23}\}$ and CSCMs $E = \{e_1,\dots,e_{16}\}$\\

\Indp
\ForEach{$\textnormal{v} \in \textnormal{V}$}{
    $d(v) \gets \sum_{e \in E} H(v,e)$
}
\Indm

\Indp
\ForEach{$\textnormal{e} \in \textnormal{E}$}{
    $\delta(e) \gets \sum_{v \in V} H(v,e)$
}
\Indm

\vspace{0.5em}
\textbf{\textit{Step 2: Build transition matrices from hypergraph}}

$P_{ue} \gets D_v^{-1} \cdot H$\\
$P_{ev} \gets D_e^{-1} \cdot H^T$\\
$P_{uv} \gets P_{ue} \cdot P_{ev}$\\
$P_{ce} \gets P_{ev} \cdot P_{ve}$

\vspace{0.5em}
\textbf{\textit{Step 3: Compute stationary distributions}}

Solve $\pi_v$ such that $\pi_v \cdot P_{uv} = \pi_v$\\
Solve $\pi_e$ such that $\pi_e \cdot P_{ce} = \pi_e$

\vspace{0.5em}
\textbf{\textit{Step 4: Compute mutual information}}

$Z \gets \sum_{v,e} H(v,e)$\\

\Indp
\ForEach{$\textnormal{v} \in \textnormal{V}$}{
    $p_v(v) \gets \sum_e H(v,e)/Z$
}
\Indm

\Indp
\ForEach{$\textnormal{e} \in \textnormal{E}$}{
    $p_e(e) \gets \sum_v H(v,e)/Z$
}
\Indm

\Indp
\ForEach{$(\textnormal{v},\textnormal{e})$ such that $H(v,e) = 1$}{
    $p_{ve}(v,e) \gets H(v,e)/Z$
}
\Indm

\Indp
\ForEach{$\textnormal{v} \in \textnormal{V}$}{
    $I(v) \gets \sum_{e \in E} p_{ve}(v,e) \log_2 \left( \frac{p_{ve}(v,e)}{p_v(v) p_e(e)} \right)$
}
\Indm

\Indp
\ForEach{$\textnormal{e} \in \textnormal{E}$}{
    $I(e) \gets \sum_{v \in V} p_{ve}(v,e) \log_2 \left( \frac{p_{ve}(v,e)}{p_v(v) p_e(e)} \right)$
}
\Indm
\vspace{0.5em}

\end{algorithm}

\end{small}

\clearpage 
\bibliographystyle{unsrt.bst}

\end{document}